\documentclass[%
 reprint,
 superscriptaddress,
 amsmath,amssymb,
 aps,
 prl,
 nolongbibliography,
]{revtex4-2}

\usepackage{dcolumn}
\usepackage{bm}
\usepackage{amsfonts,amsmath,amssymb}
\usepackage{float}
\usepackage[normalem]{ulem}
\usepackage[utf8]{inputenc}
\usepackage{graphicx}
\usepackage[a4paper,margin=1in]{geometry}
\usepackage{hyperref}
\usepackage[version=4]{mhchem}
\usepackage{xcolor}

\begin{document}

\preprint{APS/123-QED}

\title{Saturable time-varying mirror based on an ENZ material}

\author{Romain Tirole}
\affiliation{The Blackett Laboratory, Department of Physics, Imperial College London, London SW7 2BW, United Kingdom}

\author{Emanuele Galiffi}
\affiliation{The Blackett Laboratory, Department of Physics, Imperial College London, London SW7 2BW, United Kingdom}
\affiliation{Photonics Initiative, Advanced Science Research Center, City University of New York, 85 St. Nicholas Terrace, 10031, New York, NY, USA}

\author{Jakub Dranczewski}
\affiliation{The Blackett Laboratory, Department of Physics, Imperial College London, London SW7 2BW, United Kingdom}

\author{Taran Attavar}
\affiliation{The Blackett Laboratory, Department of Physics, Imperial College London, London SW7 2BW, United Kingdom}

\author{Benjamin Tilmann}
\affiliation{Chair in Hybrid Nanosystems, Nanoinstitut München, Ludwig-Maximilians-Universität München, 80539 München, Germany}

\author{Yao-Ting Wang}
\affiliation{Photonics Initiative, Advanced Science Research Center, City University of New York, 85 St. Nicholas Terrace, 10031, New York, NY, USA}

\author{Paloma A. Huidobro}
\affiliation{Instituto de Telecomunica\c c\~oes, Instituto Superior Tecnico-University of Lisbon, Avenida Rovisco Pais 1, Lisboa, 1049-001 Portugal}

\author{Andrea Al\'u}
\affiliation{Photonics Initiative, Advanced Science Research Center, City University of New York, 85 St. Nicholas Terrace, 10031, New York, NY, USA}
\affiliation{Physics Program, Graduate Center, City University of New York, New York, NY 10016, USA}

\author{John B. Pendry}
\affiliation{The Blackett Laboratory, Department of Physics, Imperial College London, London SW7 2BW, United Kingdom}

\author{Stefan A. Maier}
\affiliation{The Blackett Laboratory, Department of Physics, Imperial College London, London SW7 2BW, United Kingdom}
\affiliation{Chair in Hybrid Nanosystems, Nanoinstitut München, Ludwig-Maximilians-Universität München, 80539 München, Germany}

\author{Stefano Vezzoli}
\affiliation{The Blackett Laboratory, Department of Physics, Imperial College London, London SW7 2BW, United Kingdom}

\author{Riccardo Sapienza}
\affiliation{The Blackett Laboratory, Department of Physics, Imperial College London, London SW7 2BW, United Kingdom}

\email{romain.tirole16@imperial.ac.uk,r.sapienza@imperial.ac.uk}
    
    
\date{\today}

\begin{abstract}

We report a switchable time-varying mirror, composed of an ITO-Au bilayer,  displaying a ten-fold modulation of reflectivity ($\Delta R \approx 0.6$), which saturates for a driving pump intensity $I_{\mathrm{pump}}\approx 100$~GW/cm$^2$. Upon interacting with the saturated time-varying mirror, the frequency content of a reflected pulse is extended up to 31~THz, well beyond the pump spectral content (2.8 THz). We interpret the spectral broadening as a progressive shortening of the mirror rise time from 110 fs to sub 30 fs with increasing pump power, which is confirmed by four-wave mixing experiments  and partially captured by a linear time-varying model of the mirror. A temporal response unbounded by the pump bandwidth opens new avenues for spectral manipulation from time-varying systems with impact for communication networks, optical switching and computing.

\end{abstract}

\maketitle


Fundamental wave phenomena such as refraction and reflection rely on altering the wavevector of a wave by spatially modulating the permittivity, e.g. in a metamaterial. Conversely, a structured permittivity in the time domain manipulates the wave frequency. Time-varying systems are a platform for a wealth of exotic physics~\cite{galiffi2021photonics}, including nonreciprocity~\cite{sounas2017non,Koutserimpas2018PRLNonreciprocal,huidobro2019fresnel}, amplification \cite{galiffi2019broadband,braidotti2020zel} and topology~\cite{lin2016photonic,Lustig2018OpticaTopological,he2019floquet,darabi2020reconfigurable}, as well as quantum-relativistic effects~\cite{Nation2012RevModPhyColloquium,Braidotti2020PRLPenrose}, promising novel applications including photonic refrigeration~\cite{Buddhiraju2020PRLPhotonic} and temporal aiming \cite{pacheco2020temporal}. However, bridging the gap between theory and experiment in time-varying systems has proven especially challenging at optical frequencies and ultrafast time-scales. Until recently, optical experiments have been mostly confined to silicon waveguides~\cite{Lira2012PRLElectrically}, micro-ring resonators~\cite{Preble2007NatPhotChanging} or slow-light photonic crystals~\cite{Kampfrath2010PRAUltrafast}, but at the cost of a narrow bandwidth and a slow temporal response.

In contrast, nonlinear optical interactions in epsilon-near-zero (ENZ) materials, such as transparent conducting oxides (TCOs)~\cite{Kinsey2019NatRevMatNear,Alam2016ScienceLarge,Caspani2016PRLEnhanced,Luk2015APLEnhanced,Carnemolla2018OptMatExpDegenerate}, as Indium-Tin-Oxide (ITO), can induce modulations of the permittivity on sub-picosecond time scales~\cite{Pang2021NanoLettAdiabatic} with order-of-unity refractive index changes~\cite{Alam2016ScienceLarge}.

Time-varying physics have been observed in TCOs, with effects such as time-refraction~\cite{Zhou2020NatCommBroadband,Liu2021OptLettTunable,Bruno2020AppSciBroad, Ferrera2018JofOptUltrafast} and time-reversal~\cite{Vezzoli2018PRLOptical} of pulses, as well as in hybrid metal-dielectric metasurfaces with strong-ultrafast switching~\cite{Bohn2021NatCommAll,Taghinejad2018AdvMatHot}, frequency shifting~\cite{Alam2018NatPhotLarge,Bruno2020AppSciBroad,Pang2021NanoLettAdiabatic,Liu2021ACSPhotPhoton,Bohn2021OpticaSpatiotemporal}, polarisation switching~\cite{Taghinejad2018NanoLettUltrafast} and high-harmonic generation~\cite{Yang2019NatPhysHigh}.
Frequency shifts as large as 11~THz were achieved 
in a bulk 610~nm thick film of ITO \cite{Zhou2020NatCommBroadband}. However, the translation to thinner films, below 100~nm, has proven challenging due to the lack of interaction volume and propagation length within the active medium, with shifts limited to 1-2~THz \cite{Liu2021OptLettTunable,Bohn2021OpticaSpatiotemporal,Pang2021NanoLettAdiabatic}.
Time-varying effects have been boosted by exploiting optical resonances both in all-dielectric \cite{Shcherbakov2015NanoLettUltrafast,Shcherbakov2017NatCommUltrafast,Shcherbakov2019NatCommPhoton,Karl2020NanoLettFrequency} and hybrid plasmonic-ENZ metasurfaces~\cite{Bruno2020PRLNegative,Liu2021ACSPhotPhoton,Pang2021NanoLettAdiabatic}, albeit limited by damage thresholds lower than in bulk ITO. Large modulation amplitudes, with no spectral dynamics study, have been achieved by exploiting the Berreman resonance in GaN\cite{Dunkelberger2019ACSPhotUltrafast}.

Here, we exploit the Berreman absorption resonance of a layered ITO-gold bilayer as a switchable time-varying mirror. We demonstrate experimentally and model theoretically an ultrafast modulation of its reflection coefficient, which leads to an absolute reflectivity change $\Delta R \approx 0.6$ ($\Delta R/R\approx 1000 \%$), which saturates for a pump intensity $I_{\mathrm{pump}}\approx 100$~GW/cm$^2$. For optical pumping beyond saturation we observe the generation of new frequencies in the reflected probe spectrum which extend to a 31~THz range. 
We interpret this as a shortening of the response time of the time-varying mirror, which we confirm with a four-wave-mixing experiment.\\

\begin{figure}
    \centering
    \includegraphics[width=\columnwidth]{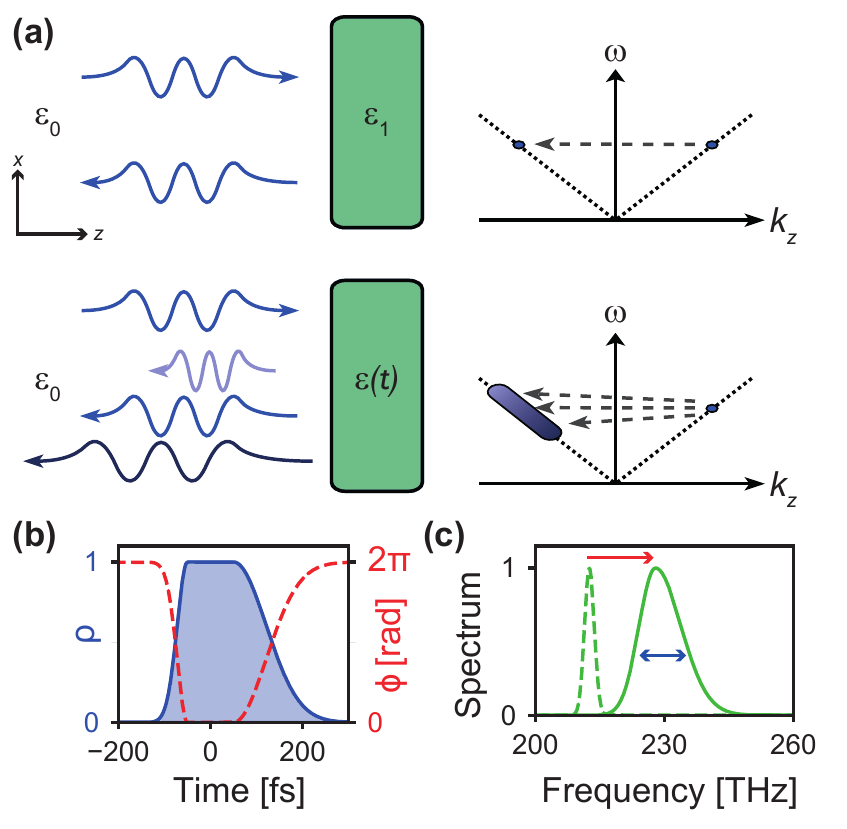}
    \caption{\textbf{(a)} Concept of a purely time-varying mirror: reflection is a transition from $k_z$ to $-k_z$ for the static mirror (top right), while the frequency distribution broadens for the time-varying mirror (bottom right). \textbf{(b)} Diagram of the temporal evolution of the complex reflection coefficient in amplitude (blue) and phase (red). \textbf{(c)} Spectral evolution of a pulse reflected by the time-varying mirror: an incoming pulse (dashed green) is shifted (red arrow) and broadened (blue arrow) in frequency (continuous green) due the respective changes in phase and amplitude of the reflection coefficient.}
    \label{fig:fig1}
\end{figure}

In a subwavelength and ultrafast time-varying mirror, as depicted in Fig.\ref{fig:fig1}(a), an incident light pulse is reflected with a time-varying complex reflection coefficient $r(t) = \rho(t) \exp[i \phi(t)]$. 
As illustrated qualitatively in Fig.\ref{fig:fig1}(b-c), a quick decrease of the Fresnel coefficient amplitude $\rho(t)$ leads to a temporal narrowing and consequently to a spectral broadening of the reflected pulse, while a change of phase $\phi(t)$ leads to a spectral shift. 

An ideal time-varying mirror should be able to provide a large change in its complex reflection coefficient, close to unity in amplitude, so that it can be completely switched on and off, and with a phase shift of order $\pi$ radians, over time-scales of the order of the optical period ($\sim$fs), in order to give rise to significant spectral modulation. 

\begin{figure*}[htb]
    \centering
    \includegraphics[width=14cm]{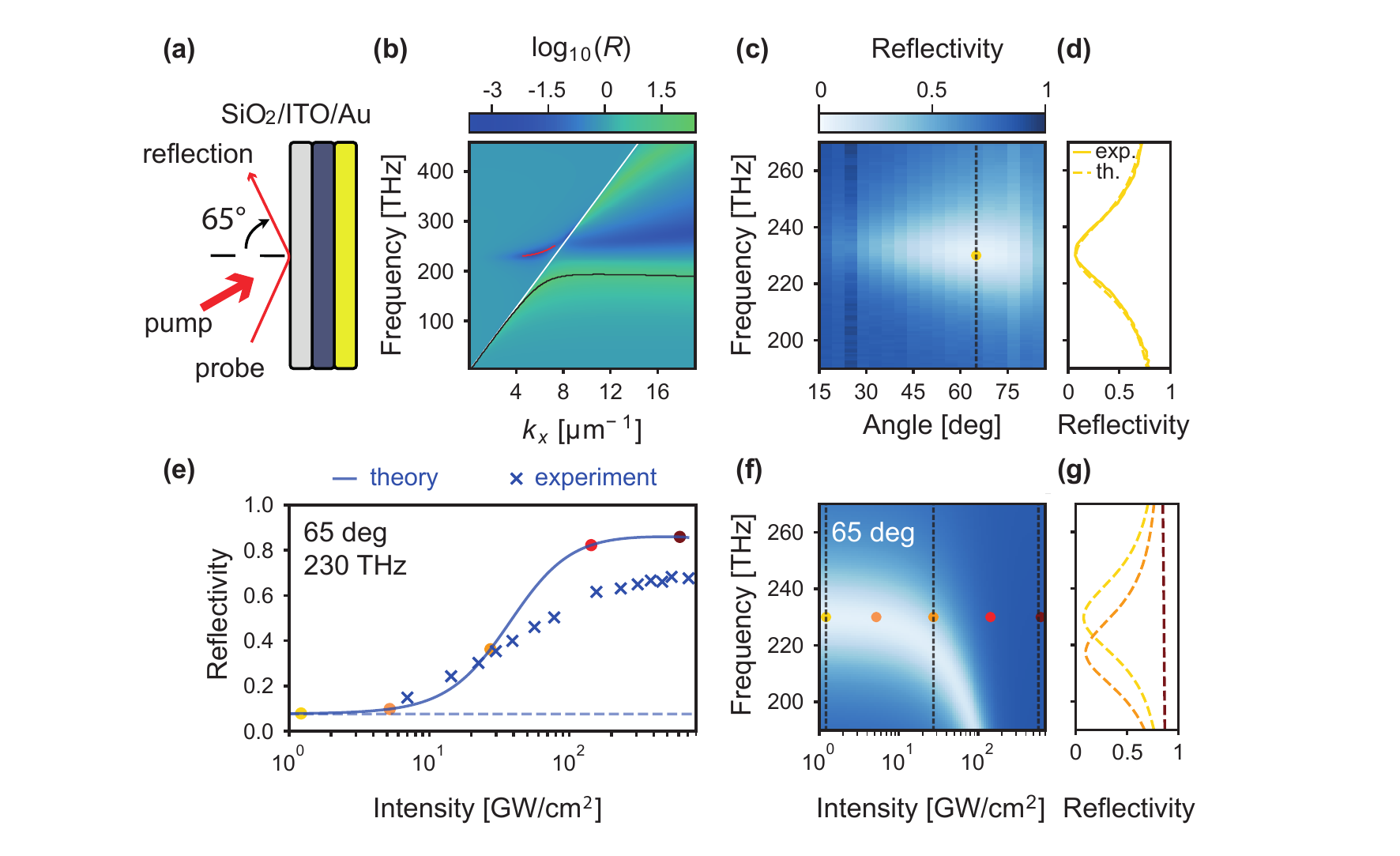}
    \caption{\textbf{(a)} Sketch of a pump-probe experiment. \textbf{(b)} Dispersion diagram of the Berreman mode (red line), situated above the light line (white line), and of the plasmonic ENZ mode (black line). \textbf{(c)} Experimental reflectivity of the sample, versus angle and frequency. The black dashed line indicates the angle of 65° at which the reflectivity in panel \textbf{(d)} is shown. \textbf{(d)} Experimental (continuous blue) and numerically obtained (dotted blue) reflectivity of the sample at an incidence angle of 65°.  \textbf{(e)} Experimental and expected reflectivity at 230 THz as a function of pump intensity. The coloured dots identify the reflectivity for a few powers as in (f). \textbf{(f)} Simulated shift of the Berreman resonance spectrum as a function of pump intensity for an illumination angle of 65°. The dashed grey lines indicate the intensities at which the spectra in panel (g) are shown. \textbf{(g)} Simulated reflectivity spectra of the sample under incident intensities of 1, 27 and 613~GW/cm$^2$.}
    \label{fig:fig2}
\end{figure*}

We study a subwavelength time-varying mirror made of a 40~nm thin film of ITO with ENZ frequency $f_{\mathrm{ENZ}} \approx$ 227~THz, deposited on glass and covered by a 100~nm gold layer as depicted in Fig.\ref{fig:fig2}(a). An optical pump pulse is used to modulate the mirror reflectivity, whereas a probe beam is used to measure the induced changes. 
The structure is much thinner than the effective wavelength in the layer ($\sim6.6$~\textmu m at $f_{\mathrm{ENZ}}$).
Thin films of ENZ materials exhibit plasmonic ENZ modes (black line in Fig.\ref{fig:fig2}(b)) beyond the light cone, as well as a Berreman mode above the light line (red-marked reflectance dip (see Supplementary Material (SM) for details)~\cite{Vassant2012OptExpBerreman}. The deeply-subwavelength thickness of the system allows efficient coupling to the resonant guided Berreman mode from free space, while the reflective gold layer increases the confinement of the mode into the ITO film, enabling strong nonlinear interactions\cite{Yang2019NatPhysHigh}.
As shown in Fig.\ref{fig:fig2}(c), experimental evidence of the efficient coupling to the Berreman mode is manifested in the strong reflectivity dip measured at an angle of 65° close to the Brewster angle, for frequencies just below $f_{\mathrm{ENZ}}$ (230~THz, 1300~nm), which we choose as the working frequency for both the pump and the probe beam in our time-varying experiment. This compares very well to the theoretical prediction shown in Fig.\ref{fig:fig2}(d).
The broad angular dependence of the Berreman mode ($\approx$ 60° to 70°) allows for the resonant coupling of both pump and probe beams. 

\begin{figure}
    \centering
    \includegraphics[width=\columnwidth]{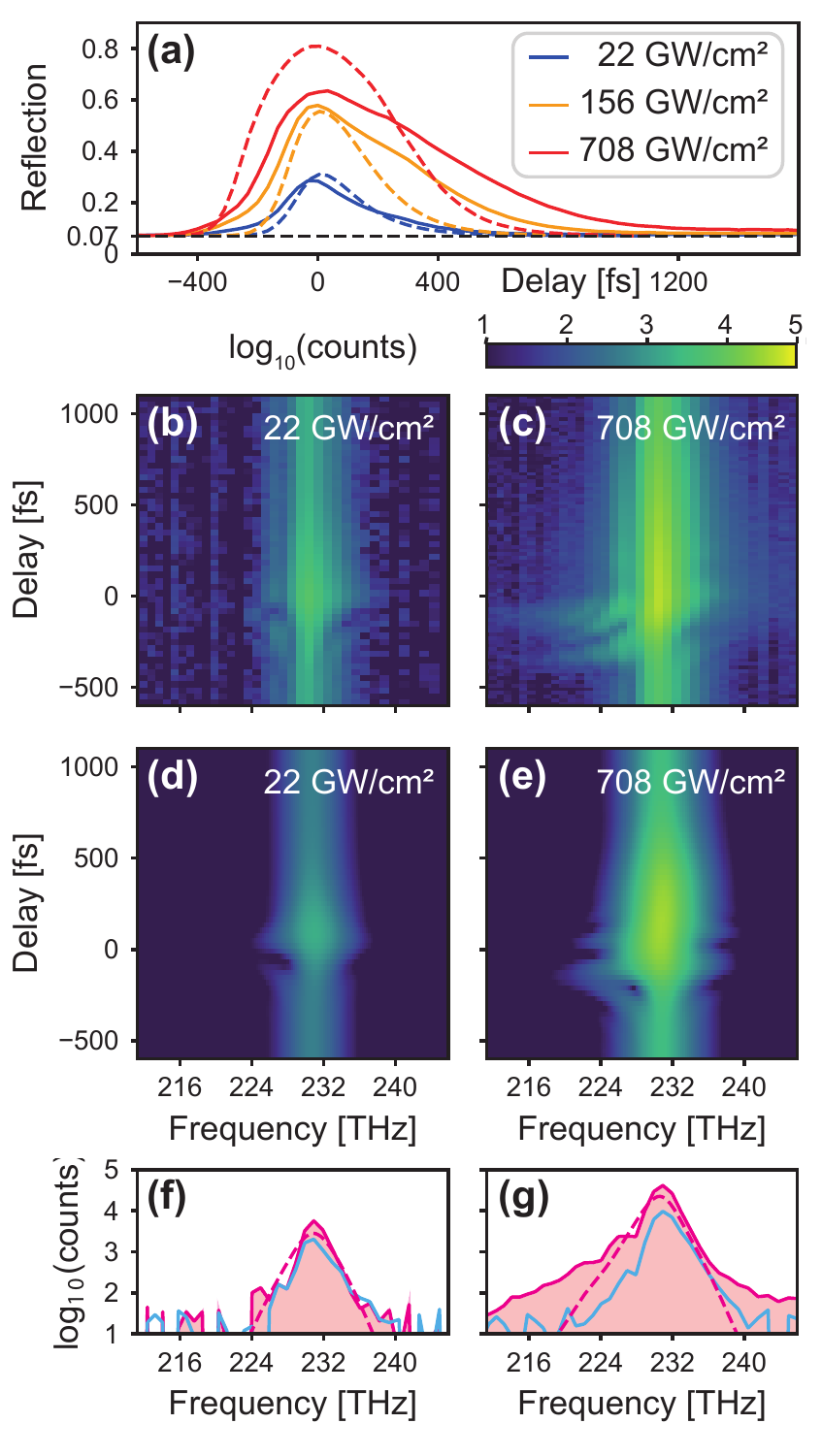}
    \caption{\textbf{(a)} Experimental (continuous line) and numerically simulated (dashed line) evolution of the probe reflection as a function of the pump/probe delay for three pump intensities. \textbf{(b-e)} Comparison of the reflection as a function of delay and frequency, on a log scale, at low intensity (22~GW/cm$^2$, (b) experiment, (d) theory) and at high intensity (708~GW/cm$^2$, (c) experiment, (e) theory). \textbf{(f,g)} Cross-section of the experimental reflection signal shown in (b-c), at a delay of $-90$~fs (pink line) as well as -500~fs delay (light blue line) where no modulation happens. Numerically simulated modulated spectra from (d,e) are shown in dashed pink.}
    \label{fig:fig3}
\end{figure}

The reflectivity of the ITO/Au mirror can be efficiently modulated by an ultrafast laser pulse, because of the consequent reduction in the plasma frequency following a change of the photocarrier dynamics in the ITO layer \cite{Taghinejad2018AdvMatHot,Wang2019PRAExtended}. The plasma frequency $f_p$ shift is assumed to be proportional to the pump intensity \cite{Bohn2021OpticaSpatiotemporal} in our model (see SM). This reduction in $f_p$ is due to pump-induced intraband electronic transitions to states with higher effective mass~\cite{Wang2019PRAExtended}. From this model, we predict the evolution of the reflection modulation with increasing pump intensity, shown in Fig.\ref{fig:fig2}(e): the reflectivity is expected to reach a plateau for a driving intensity $\sim$100~GW/cm$^2$, which is confirmed by the data (blue crosses).
Pump and probe beams are degenerate at the Berreman frequency (230~THz), the pump is incident at 75° and the probe at 65°. The mirror total reflectivity changes from $R=0.07$ (without pump) and converges to $R=0.66$ at intensities above saturation ($I_{\mathrm{sat}}\approx 100$~GW/cm$^2$), corresponding to an observed modulation of $\Delta R/R\sim$1000\%, as plotted in Fig.\ref{fig:fig3}(a). When the reflectivity of the glass substrate $R_s = 0.04$, which is not modulated, is subtracted, the observe reflectivity change is as high as 2200\%. 

This saturation of the reflection modulation can be explained by the redshift of the Berreman mode with $f_p$, illustrated in Fig.\ref{fig:fig2}(f): once the Berreman reflectivity dip is shifted far enough from its original frequency, the reflectivity reaches its highest value (Fig.\ref{fig:fig2}(g)).

The role of the Berreman resonance is to increase the contrast of the maximum achievable modulation, as illustrated in Fig.\ref{fig:fig2}(e-g), and to enhance the pump intensity in the ITO layer. 
The combination of these features, together with the large damage threshold of ITO, makes the Berreman mode a convenient platform to explore time modulations in a strongly non-perturbative pumping regime, i.e. well above the saturation of the reflectivity modulation. \\

So far, we have discussed the maximum achievable modulation and the mechanism leading to the saturation of reflectivity. We now turn to discuss the time dynamics of the mirror. The reflectivity changes are driven by pump pulse of finite duration (220 fs and 2.8~THZ, or 16~nm, bandwidth) and probe by an identical pulse. The evolution of the probe reflectivity with the pump/probe delay is illustrated in Fig.\ref{fig:fig3}(a) for different pump powers.  
The reflectivity rise time is given by the convolution between the pump pulse and the faster material response of ITO (below 50~fs~\cite{Pang2021NanoLettAdiabatic}, set at 30~fs in our model). The reflectivity decay time, on the other hand, is dominated by the slower relaxation dynamics of ITO ($\approx 360$ fs \cite{Alam2016ScienceLarge}).
Beyond $I_{\mathrm{sat}}$, after a sharp rise, the reflectivity reaches a plateau and features a longer decay time (yellow and red lines in Fig.\ref{fig:fig3}(a)). Our numerical simulations reproduce the temporal dynamics with good accuracy at low pump intensity (dashed blue line in Fig.\ref{fig:fig3}(a)), but overshoot in amplitude  and fail to capture the longer decay time and the flattening of the response at high pump intensities. 

In order to characterise the time-varying effects of the mirror, we focus on the spectral features of the probe signal undergoing modulation. Fig.\ref{fig:fig3} shows the measured (b) and simulated (d) evolution of the probe spectrum at various delays, for a pump power of $I_{\mathrm{pump}}=22$~GW/cm$^2$, well below saturation. 
A successive thinning and broadening in the spectrum around zero delay is observable in Fig.\ref{fig:fig3}(b), associated with the amplitude change of the reflection coefficient. Overall the probe spectral content is largely unchanged, as shown in Fig.\ref{fig:fig3}(f) where the modulated (pink line) and unmodulated (light blue line) spectra are almost superposed.

These below-saturation dynamics, including the fine features around zero-delay, are well reproduced by the linear time-varying model shown in Fig.\ref{fig:fig3}(d). Since the modulation is significantly faster during the rise time, the time evolution of the spectrum is not symmetric and neither is the amount of blue and red shift at different delays.
 
In Fig.\ref{fig:fig3}(c) we present the spectral map for the maximum pump intensity achievable in our setup, $I_{\mathrm{pump}}=708$~GW/cm$^2$, well beyond $I_{\mathrm{sat}}$. New frequencies are generated beyond the bandwidth of the unmodulated spectrum.
This is qualitatively similar to what is predicted by our model and plotted in Fig.\ref{fig:fig3}(e) and in Fig.\ref{fig:fig3}(g) where new frequencies are observed (pink line) well beyond the frequency content of the unmodulated case (light blue line). The broadening and shift of the central part of the spectrum is well captured by the time-varying model (pink dashed line), however tails are much larger than predicted. Fringes extended from 214 to 245~THz appear in the map of Fig.\ref{fig:fig3}(c), well beyond the bandwidth ($2.8$~THz) of the modulating pump pulse. The spectral broadening is about 31~THz at an intensity of $10^{-3}$ of the modulated pulse, 10x higher than the noise level and it is largest at slightly negative delay ($-90$~fs), on the rise time of the mirror reflectivity, when the temporal gradient of reflectivity is at its maximum. 
Moreover, the spectrum becomes visibly asymmetric, with a clear inclination towards the red, due to change of phase in time. The FWHM on the spectrum, however, is only marginally affected with an increase of 0.2~THz and a maximum recorded shift of 0.3~THz. The low amplitude of the newly generated frequencies can be explained by the probe pulse, which is longer than the medium rise time, and therefore only partially modulated. 
 Our simple time varying model predicts a shortening of rise time when saturating the mirror reflection coefficient, going from the original 110 fs (pump limited) to about 60 fs. However, the mismatch between the data and the model in the tails of Fig.\ref{fig:fig3}(g) suggests that the rise time could be much shorter, pointing to richer material dynamics emerging under saturation conditions.\\

The acceleration of the response time with increasing pump intensity can be measured more precisely with a four-wave-mixing (FWM) experiment. FWM is only generated when pump and probe beams interact inside the ITO layer, during the switching on of the time-varying mirror.  . 

\begin{figure}
    \centering
    \includegraphics[width=\columnwidth]{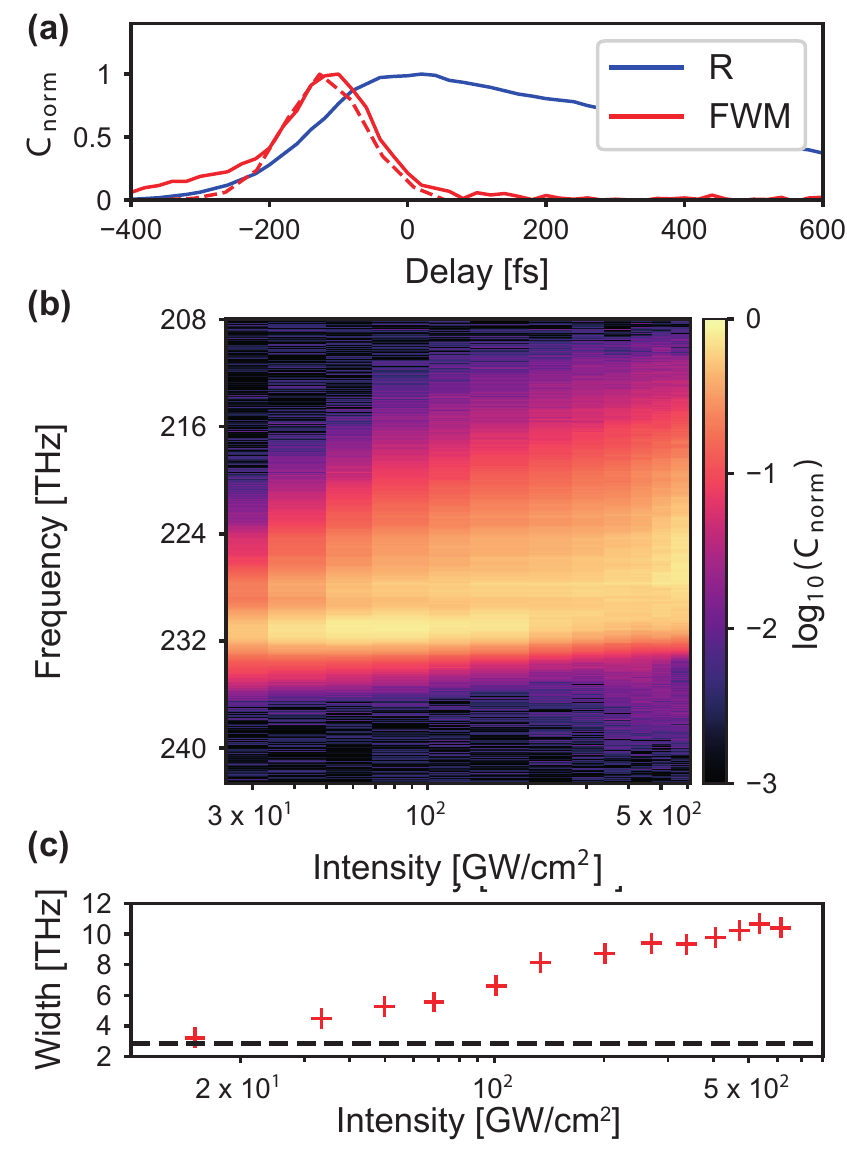}
    \caption{\textbf{(a)} Measured evolution with pump/probe delay of the FWM signal (continuous red line) versus reflection (blue line). Simulated FWM signal is shown by the red dashed line. \textbf{(b)} Measured evolution of the FWM spectrum as a function of pump intensity: the spectrum gets broader with pump power, reaching a maximum bandwidth of 10.7~THz (61.5~nm) and a redshift of 3.8 THz (22~nm) for a pump intensity of 535~GW/cm$^2$. \textbf{(c)} Measured broadening of the FWM signal as a function of pump intensity (red crosses), compared with the spectral width of the original input probe spectrum (dashed black line).}
    \label{fig:fig4}
\end{figure}

The measured FWM signal as a function of pump/probe delay is shown in Fig.\ref{fig:fig4}(a) (solid red line). As expected, FWM generation mainly occurs during the rise time of the reflection (blue line in Fig.\ref{fig:fig4}(a)), as also confirmed by numerical simulation (dashed red line, see SM). A high FWM efficiency of 7.5\%, defined as the ratio of FWM power to the input probe power, is observed for a pump intensity of 612~GW/cm$^2$~(see Fig. S4 in SM). 
In Fig.\ref{fig:fig4}(b) we plot the FWM spectrum as a function of the pump intensity for a delay of about $-150$~fs, where the FWM signal is maximum. The spectrum starts from a nearly Gaussian shape centered at 231~THz and with a width of 3.2~THz at a pump intensity of 15~GW/cm$^2$, very close to the original probe spectrum (2.8~THz). This spectrum evolves towards a broad distribution with a FWHM of 10.7~THz (61.5~nm) for pump powers of $\sim$600~GW/cm$^2$, and which is also red-shifted by about 3.8~THz.

The width of the FWM signal shown in Fig.\ref{fig:fig4}(b) is plotted in Fig.\ref{fig:fig4}(c). The spectral width is taken here at the FWHM, as the time modulation affects the whole FWM spectrum, unlike the probe reflection. Therefore, from this 3.8x increase in bandwidth, we can estimate that the rise time of the time-varying mirror shortens to $\sim30$~fs. The FWM spectral broadening slows down around $\sim$400~GW/cm$^2$, as the medium reaches its intrinsic rise time. The value of 30~fs estimated from FWM data is likely an overestimate, as FWM is a complex process and other effects come into play in determining its bandwidth. \\





In conclusion, we have presented a saturable time-varying mirror based on the Berreman resonance of an ITO-Au thin film, exhibiting a record modulation of over 1000\% in reflectivity.
When pumping above the saturation level of the mirror response, new frequencies are generated, as far as 5 bandwidths away from the original carrier frequency, as confirmed by FWM measurements. 
We interpret the spectral broadening as a progressive acceleration of the mirror response time, which can be controlled by simply adjusting the pump pulse intensity, and not its duration, when in the non-perturbative regime 
~\cite{Yang2019NatPhysHigh,Bruno2020AppSciBroad}. Although the Berreman resonance of the ITO/Au bilayer allows access to the saturation regime for a moderate pump intensity of 100~GW/cm$^2$, our results are very general and similar dynamics could emerge in other resonant architectures, once the strong saturation of their response is reached.
The extent of the generated frequencies in both the reflection and the FWM spectra, pointing at a shortening of the rise time beyond the adiabatic limit, calls for a further development of the theory of photocarrier dynamics in ITO and ENZ media in the non-adiabatic regime
and for a deeper understanding of the physical origin of saturation.

The results described here show the potential for efficient spectral manipulation of light, unlocking applications such as pulse shaping and temporal holograms. They represent a firm step towards the experimental realisation of fundamental concepts like time crystals, time diffraction and spatiotemporal metasurfaces. 

\begin{acknowledgments}

R.S., J.P., and S.V. acknowledge funding from the Engineering and Physical Sciences Research Council (EP/V048880), J.P. from the Gordon and Betty More Foundation, E.G.  from the EPSRC (EP/T51780X/1) and the Simons Society of Fellows (855344,EG), A.A. the Department of Defense, the Simons Foundation, and the AFOSR MURI program, P.A.H. from FCT (UIDB/50008/2020 and CEECIND/02947/2020). 


%
Data and codes are available upon request.

\end{acknowledgments}


\renewcommand{\section}[2]{}
\bibliography{main}

\end{document}


\preprint{APS/123-QED}

\title{Supplementary Material: Saturable time-varying mirror based on an ENZ material}

\author{Romain Tirole}
\affiliation{The Blackett Laboratory, Department of Physics, Imperial College London, London SW7 2BW, United Kingdom}

\author{Emanuele Galiffi}
\affiliation{The Blackett Laboratory, Department of Physics, Imperial College London, London SW7 2BW, United Kingdom}
\affiliation{Photonics Initiative, Advanced Science Research Center, City University of New York, 85 St. Nicholas Terrace, 10031, New York, NY, USA}

\author{Jakub Dranczewski}
\affiliation{The Blackett Laboratory, Department of Physics, Imperial College London, London SW7 2BW, United Kingdom}

\author{Taran Attavar}
\affiliation{The Blackett Laboratory, Department of Physics, Imperial College London, London SW7 2BW, United Kingdom}

\author{Benjamin Tilmann}
\affiliation{Chair in Hybrid Nanosystems, Nanoinstitut München, Ludwig-Maximilians-Universität München, 80539 München, Germany}

\author{Yao-Ting Wang}
\affiliation{Photonics Initiative, Advanced Science Research Center, City University of New York, 85 St. Nicholas Terrace, 10031, New York, NY, USA}

\author{Paloma A. Huidobro}
\affiliation{Instituto de Telecomunica\c c\~oes, Instituto Superior Tecnico-University of Lisbon, Avenida Rovisco Pais 1, Lisboa, 1049-001 Portugal}

\author{Andrea Al\'u}
\affiliation{Photonics Initiative, Advanced Science Research Center, City University of New York, 85 St. Nicholas Terrace, 10031, New York, NY, USA}
\affiliation{Physics Program, Graduate Center, City University of New York, New York, NY 10016, USA}

\author{John B. Pendry}
\affiliation{The Blackett Laboratory, Department of Physics, Imperial College London, London SW7 2BW, United Kingdom}

\author{Stefan A. Maier}
\affiliation{The Blackett Laboratory, Department of Physics, Imperial College London, London SW7 2BW, United Kingdom}
\affiliation{Chair in Hybrid Nanosystems, Nanoinstitut München, Ludwig-Maximilians-Universität München, 80539 München, Germany}

\author{Stefano Vezzoli}
\affiliation{The Blackett Laboratory, Department of Physics, Imperial College London, London SW7 2BW, United Kingdom}

\author{Riccardo Sapienza}
\affiliation{The Blackett Laboratory, Department of Physics, Imperial College London, London SW7 2BW, United Kingdom}

\email{romain.tirole16@imperial.ac.uk,r.sapienza@imperial.ac.uk}

\date{\today}
    
\maketitle
    
\section{Methods}


\subsection{Sample Fabrication}

A layer of 100~nm of gold was deposited on a commercially available 40~nm ITO thin film (from Präzisions Glas \& Optik GmbH) with an Angstrom Engineering Amod vacuum deposition system. The complex permittivity of the ITO layer is shown in Fig.\ref{fig:supp1}(a).


\subsection{Optical Setup}

We perform a pump-probe experiment for light incident at various angles on the sample. 220 fs optical pulses are generated from a LightConversion Pharos laser whose wavelength is varied within the NIR around 1300 nm using a LightConversion Orpheus optical parametric amplifier. The beam is split into pump and probe paths using an 80:20 beam splitter, and their respective intensities are controlled with neutral density filters. The probe is sent through a variable delay stage before being focused at the same position as the pump on the sample. The reflected probe signal and generated four-wave-mixing signal are collected and sent via fibers to a OceanOptics flameNIR spectrometer to assess the spectral content of the reflected probe  and a Princeton Instruments NIRvana spectrometer to measure FWM, thus allowing mapping of the modulation of the sample in time and in frequency. Characterisation of the Berreman mode's reflectivity was done with a white light in the same setup and configuration as the probe reflection experiment.

\subsection{Characterisation of the modulation}

\begin{figure}
    \centering
    \includegraphics{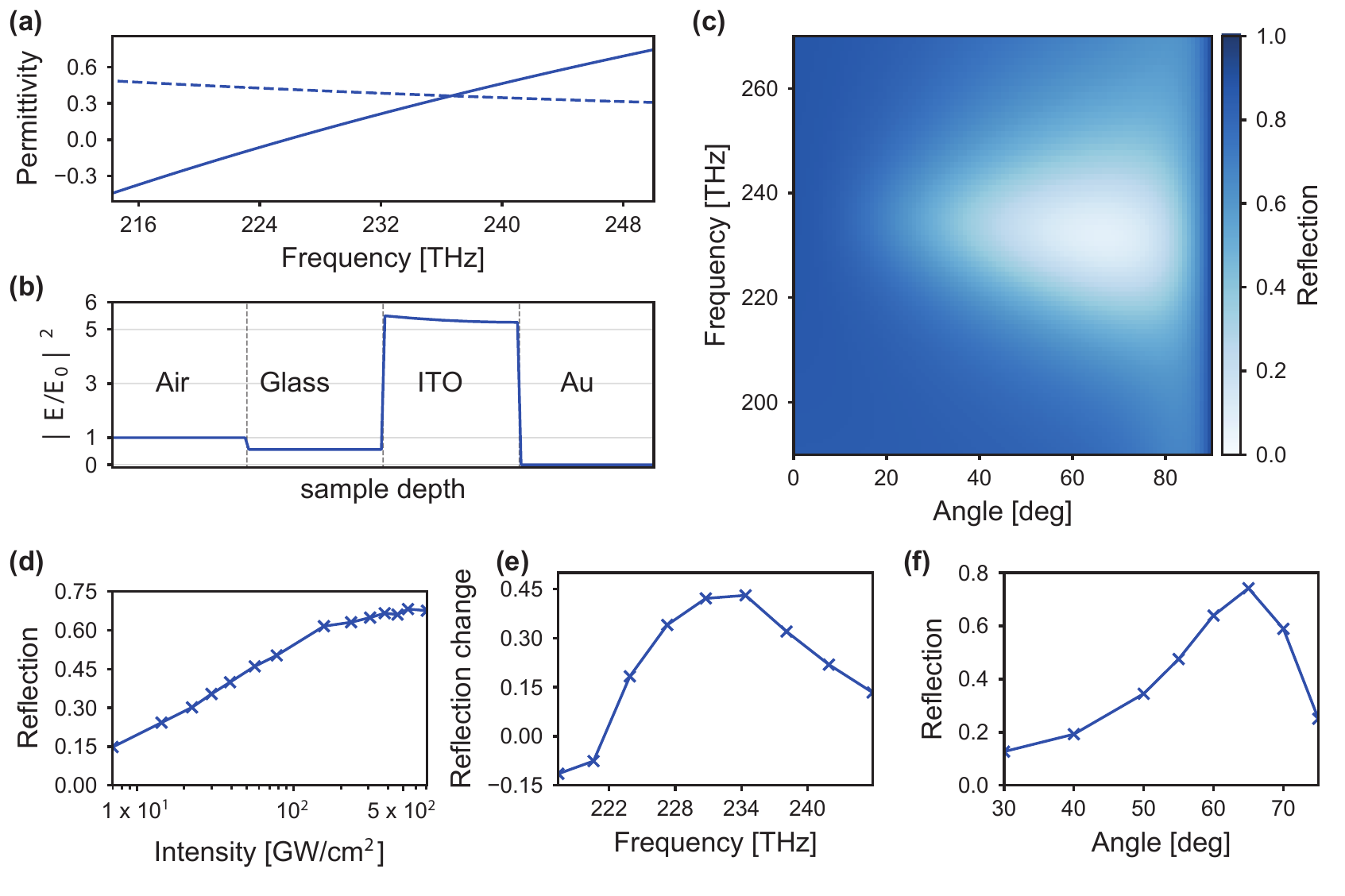}
    \caption{\textbf{(a)} Real (continuous) and imaginary (dashed) parts of the permittivity of the surface's ITO layer. \textbf{(b)} Field distribution in the different layers, computed from TMM at an angle of 65~° and a frequency of 230~THz. \textbf{(c)} Simulated linear reflection of the ITO-Au stack as a function of frequency and angle, from TMM. \textbf{(d-f)} Probe reflection signal dependence on \textbf{(d)} pump power intensity for a probe angle of 65\textdegree~and central frequency 230~THz \textbf{(e)} pump and probe central frequency for a probe angle of 65~ and a pump intensity of 29~GW/cm\textsuperscript{2} \textbf{(f)} probe angle for a central frequency of 230~THz and pump intensity of 132~GW/cm\textsuperscript{2}.}
    \label{fig:supp1}
\end{figure}

The Berreman mode can be accessed from a wide range of angles (see Fig.\ref{fig:supp1}(c)), which allows both pump and probe in a pump-probe experiment to be in  resonance. This way, both the modulation strength (pump being efficiently coupled to the medium, leading to stronger nonlinear optical effects) and the modulation depth (probe being at a minimum in the reflection spectrum in the absence of modulation) are enhanced. The field enhancement prediction from our transfer matrix method (TMM) analysis is shown in Fig.\ref{fig:supp1}(b), along with the Berreman reflection spectrum with angle in Fig.\ref{fig:supp1}(c). The saturation behaviour of the modulation at high pump intensities is shown in Fig.\ref{fig:supp1}(d). The optimal configuration was determined to correspond to a pump/probe frequency of 230~THz (1300~nm wavelength) and a probe angle of 65° from the sample plane normal as can be inferred from the frequency and angular dependence of the modulation shown in Fig.\ref{fig:supp1}(e,f). Note that the measured reflection also includes a contribution  (about 4$\%$) from the \ce{SiO2}/air interface, which increases the reflection at the Berreman condition and thus reduces the achievable modulation. The decay of the material excitation happens on two timescales, a first one of 350-600~fs due to electron-phonon interaction and a second one on the order of ms due to thermal relaxation. For this reason, we restrict the laser pulse repetition rate in our pump-probe experiment to 100~Hz, in order to allow the complete relaxation of the system back to its ground state. 

\section{Modelling of time-varying ITO metamirror}

\subsection{Analytic reflection model}

The analytic reflection coefficient for the structure (Eq. \ref{eq:reflCoeff}) was used to model the linear properties of the ITO-Au stack. The frequency-dependent reflection coefficient for p-polarized waves in our SiO2-ITO-Au interface structure can be derived analytically, and is given by:
\begin{align}
    r(\omega,k_x) &= \frac{\cos(k_2 d)( \frac{k_2 \varepsilon_3}{k_3\varepsilon_2}-\frac{k_2 \varepsilon_1}{k_1 \varepsilon_2} ) - i \sin(k_2 d) (1- \frac{k_2^2 \varepsilon_1\varepsilon_3}{k_1 k_3 \varepsilon_2^2}) }{\cos(k_2 d)( \frac{k_2 \varepsilon_3}{k_3\varepsilon_2}+\frac{k_2 \varepsilon_1}{k_1 \varepsilon_2} ) - i \sin(k_2 d) (1+ \frac{k_2^2 \varepsilon_1\varepsilon_3}{k_1 k_3 \varepsilon_2^2})}
    \label{eq:reflCoeff}
\end{align}
where $k_j = \sqrt{\varepsilon_j \omega^2/c^2-k_x^2}$, with $j=1,2,3$ is the wavevector component normal to the surface in SiO2, ITO and Au respectively, $\varepsilon_j$ are the corresponding permittivities, and $d$ is the ITO thickness. The reflection results shown in Fig.~2(b-e) in the main text and Fig. ~\ref{fig:supp1}(c) are computed from Eq. \ref{eq:reflCoeff}.

The permittivity of SiO2 is assumed to be dispersionless and equal to $\varepsilon_1 = 2.25$. The optical response of ITO ($j=2$) is described via a Drude model with plasma frequency $\omega_p = 2.88 \times 10^{15}$ rad/s, a background permittivity $\varepsilon_{\infty} = 4.08$ and a loss rate $\gamma = 0.145 \times 10^{15}$ rad/s. Similarly, the Drude parameters for Au ($j = 3$) are $\omega_{p} = 2.88 \times 10^{15}$ rad/s, $\varepsilon_{\infty} = 10.41$ and $\gamma = 0.407 \times 10^{15}$ rad/s.

\subsection{Time-varying material parameters}

In order to model the temporal dynamics induced by the pump system, we first calculate the time-varying modulation of the ITO plasma frequency $\omega_p$, 
\begin{equation}
    \omega_p(t) = (1+\delta\omega_{p}(t))\omega_{p,0}
\end{equation}
In order to calculate the modulation profile $\delta\omega_p(t)$ we use a standard approach based on convolution:
\begin{align}
    \delta\omega_{p}(t) = I_{pm} * \xi
\end{align}
of the gaussian pump input intensity profile $I_{pm}$ (outside of the ITO) with a response function:
\begin{equation}
        \xi(t)=
        \left\{ \begin{array}{lll}
            0 & t \leq 0 \\
            t/\tau_{rise} & 0<t<\tau_{rise} \\
            e^{-(t-\tau_{rise})/\tau_{dec}} & t>\tau_{rise}
        \end{array} \right.
\end{equation}
characterized by a rise-time $\tau_{rise} = 30$ fs, and a thermal decay time $\tau_{dec} = 210$ fs. The resulting modulation profile computed via the convolution is normalized, such that the maximum peak change $\max[\delta\omega_p] = \omega_{p,2}\max[I_{pm}]$ 
where $\omega_{p,2}$ is a constant phenomenological coefficient that incorporates the nonlinear response of ITO into the linear time-varying model.
We find $\omega_{p,2}\approx -0.42\%$ to best fit our experimental results. 


In addition, since our strategy exploits resonant absorption, we expect self-modulation of the pump absorption into the ITO due to the shifting of the Berreman mode to play a significant role. In order to account for this effect we estimate the absorbed pump power in the ITO (by subtracting the reflected part, see next subsection for the details on how reflection from the time-modulated structure is computed) and recalculate the resulting parameter modulation accordingly after the first iteration, which is then fed into the probe calculation described in the next subsection. 

\subsection{Scattering from time-modulated structure}

In order to model scattering in our time-modulated structure, while accounting for its strongly dispersive character, we implemented the approximate but efficient method introduced in Ref. \cite{Bohn2021NatCommAll}, which relies on the separation of timescales respectively associated with (a) the wave period $T=2\pi/\omega \approx 4$ fs, which enters the dispersive response of the system and (b) the temporal dynamics induced by the envelope of the pump beam (which is identical to that of the probe), whose timescale $\Delta_1 \approx 220$ fs is associated with the duration of the pump pulse for low intensities, but reduces and approaches the response time $\tau_{rise} \approx 30$ fs of the material at higher intensities in the saturation regime. We briefly report the key equations here for completeness and refer the reader to Ref. \cite{Bohn2021NatCommAll} for the complete derivation.

In order to implement the method, we define a double-frequency reflection coefficient \cite{Bohn2021NatCommAll} valid in the adiabatic limit:
\begin{align}
    r(\omega_1,\omega_2) = \int_{-\infty}^{\infty} r(\frac{t}{\Delta_1},\omega_2) e^{i\omega_1 t} \text{d}t,
\end{align}
where $\omega_1$ and $\omega_2$ are two frequency scales associated with the pump-induced material dynamics and the probe frequency respectively. The adiabatically time-dependent reflection coefficient $r(t/\Delta_1,\omega_2)$ is obtained by feeding the previously calculated time-varying $\omega_p$ into the reflection coefficient (Eq. \ref{eq:reflCoeff}).

\begin{figure}
    \centering
    \includegraphics{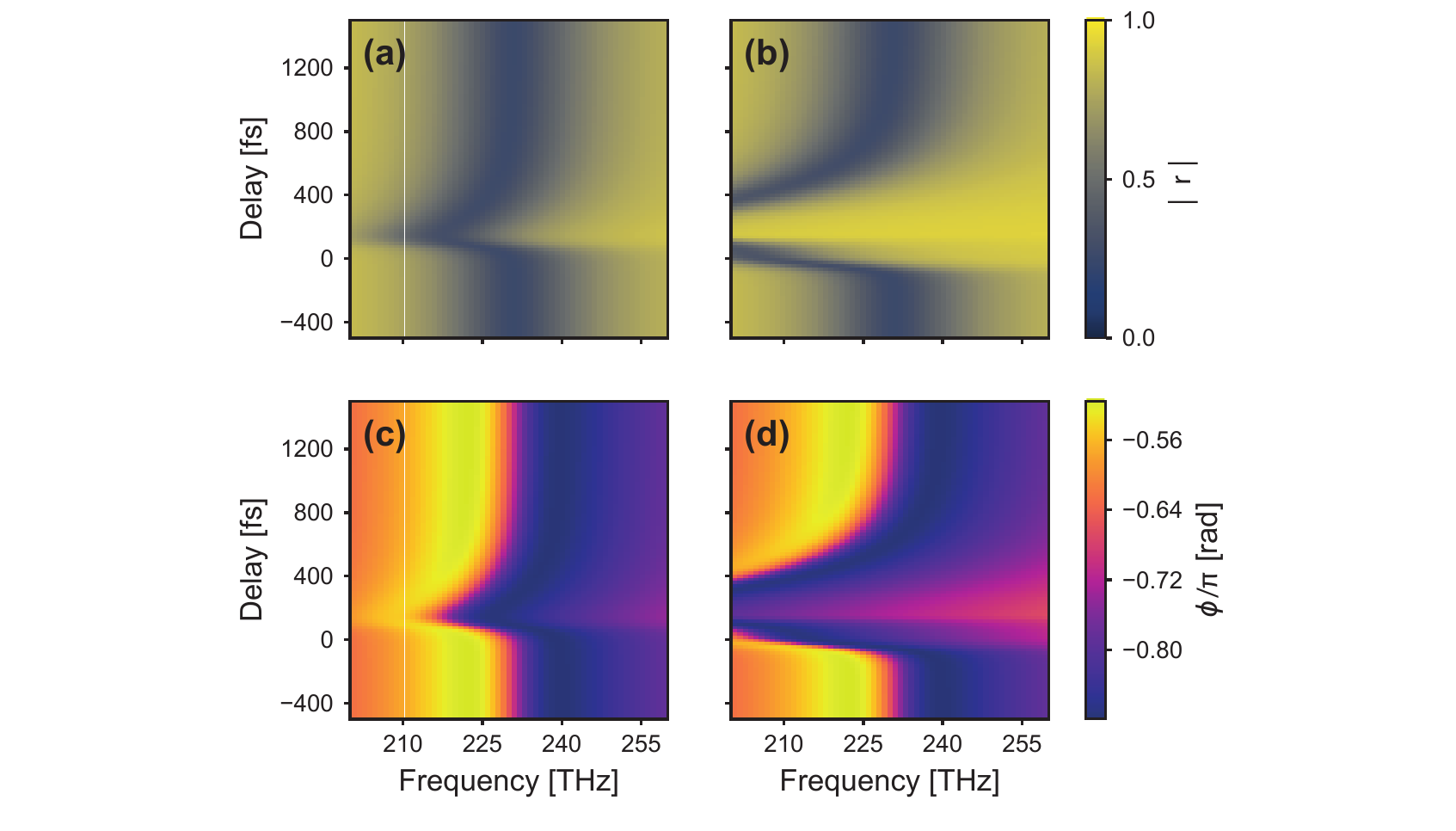}
    \caption{\textbf{(a-d)} Amplitude (a,b) and phase (c,d) of the mirror's complex reflection coefficient spectrum against delay, simulated for an intensity of 22~GW/cm\textsuperscript{2} (a,c) and 708~GW/cm\textsuperscript{2} (b,d).}
    \label{fig:supp2}
\end{figure}

The total reflected field is then given by:
\begin{align}
    H_r(\omega) &= \int_{-\infty}^{\infty} H_r(t) e^{i\omega t} dt = \int_{-\infty}^{\infty} \frac{d\omega'}{2\pi} a(\omega') r(\omega-\omega',\omega')
\end{align}
where $a(\omega')$ is the Fourier transform of the input probe pulse. The integral above may look like a convolution, but it is not, due to the presence of two frequency variables. It can be conveniently computed via two subsequent integrations:
\begin{align}
    H_r(\omega) = \int_{-\infty}^{\infty} e^{i\omega t} dt \int_{-\infty}^{\infty} \frac{d\omega'}{2\pi} a(\omega') r(t,\omega') e^{-i\omega' t},
\end{align}
the outer of which is a standard Fourier transform and can be computed via FFT, while the inner one must be computed by standard numerical integration. The probe spectrum $a(\omega)$ is a gaussian fitted to the experimental one as measured by replacing the structure with a highly reflecting silver mirror.

\subsection{Complex reflection coefficient}

The amplitude and phase of the Fresnel coefficient of the system (neglecting the air/glass interface) are shown in Fig.~\ref{fig:supp2}(a-d): (a,c) for $I=22$ GW/cm$^2$ and (b,d) for $I=708$ GW/cm$^2$. Although the frequency at the center of the Berreman resonance (230~THz) will experience the strongest change in amplitude, it is frequencies on either side that will undergo the strongest phase change and thus frequency shifts. Furthermore, the change in reflection is smaller on the edges than at the center of the Berreman resonance, and therefore we can expect a larger portion of the reflection spectrum to be modulated.


\section{Frequency shift and broadening}

\begin{figure}
    \centering
    \includegraphics{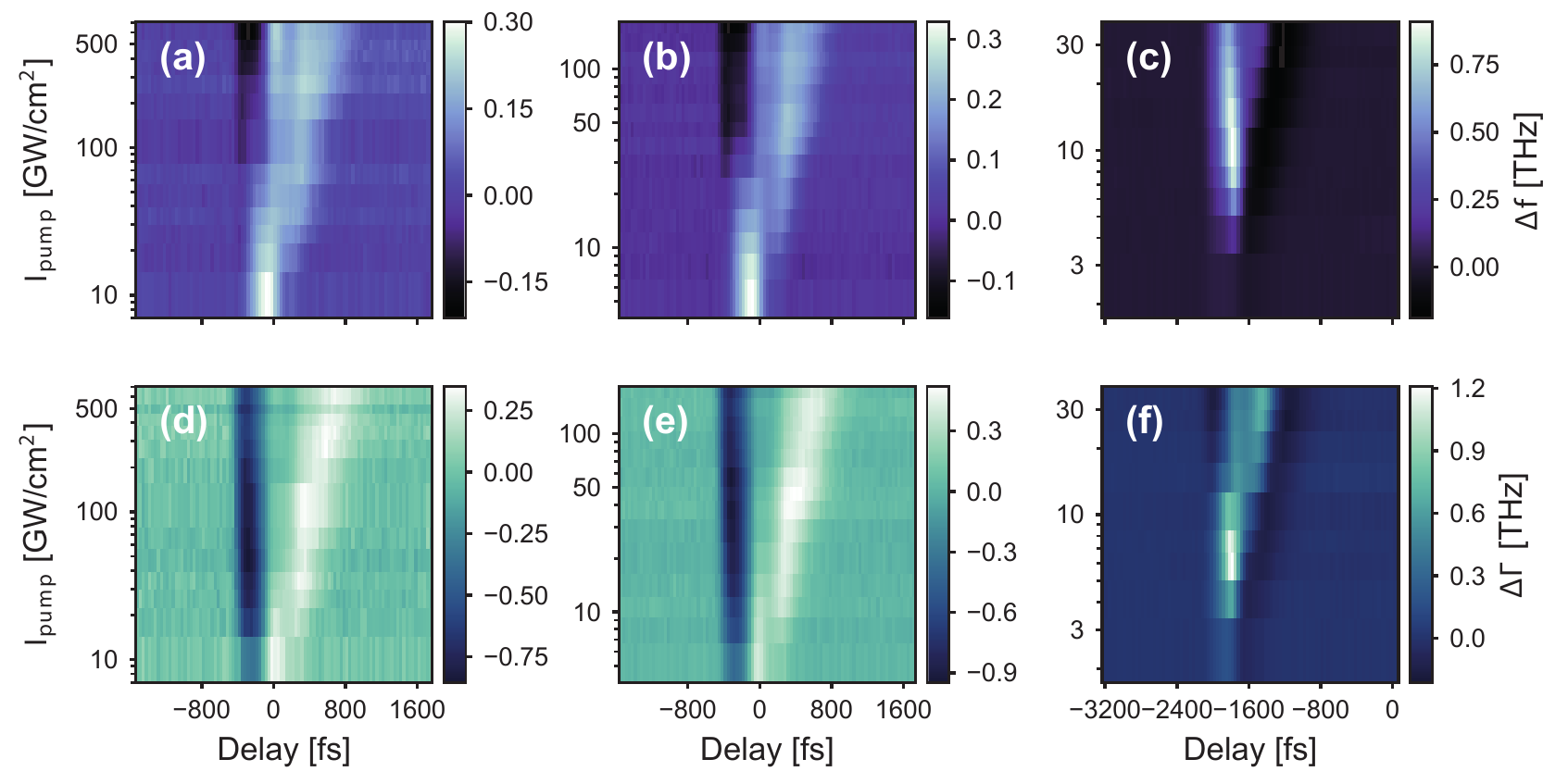}
    \caption{\textbf{(a-c)} Frequency shift and \textbf{(d-f)} broadening of the probe signal as a function of delay and pump power, with central frequency at (a,b,d,e) 230~THz and (c,f) 220~THz.}
    \label{fig:supp3}
\end{figure}

Fig.~\ref{fig:supp3} shows the shift and broadening of the spectrum with delay and pump intensity. At 230~THz, the strongest blueshifts are obtained at the lowest intensities (see Fig.\ref{fig:supp3}(a,b)). This is due to the rise in reflection of the sample, preventing the probe from entering the ITO layer and being modulated (while the phase shift changes minimally within the linear regime), coupled with the shorter interaction time between the shortening modulation and the probe at high intensity. Entering the saturation regime, a redshift appears at the rise of the modulation, that may be due to broadened frequencies undergoing a phase shift. We measured a strongest blueshift of 0.3~THz at 7~GW/cm\textsuperscript{2} for central frequency of 230~THz. In comparison, by pumping outside the Berreman resonance (much less efficiently, see Fig.\ref{fig:supp3}(c)), we measured a shift of 0.9~THz at 85~GW/cm\textsuperscript{2} for a probe centered at 220~THz. A broadening of 1.2~THz at 50~GW/cm\textsuperscript{2}  was also measured at that frequency (see Fig.\ref{fig:supp3}(f)), that we attribute to the simultaneous action of the change in reflectivity (about 0.3) coupled to the strong phase shifts in this spectral region. Strong changes in width of the spectrum at 230~THz are also observed, with a successive thinning and broadening of -0.8~THz and 0.3~THz at saturation (see Fig.\ref{fig:supp3}(d,e)).

\section{FWM efficiency}

\begin{figure}
    \centering
    \includegraphics{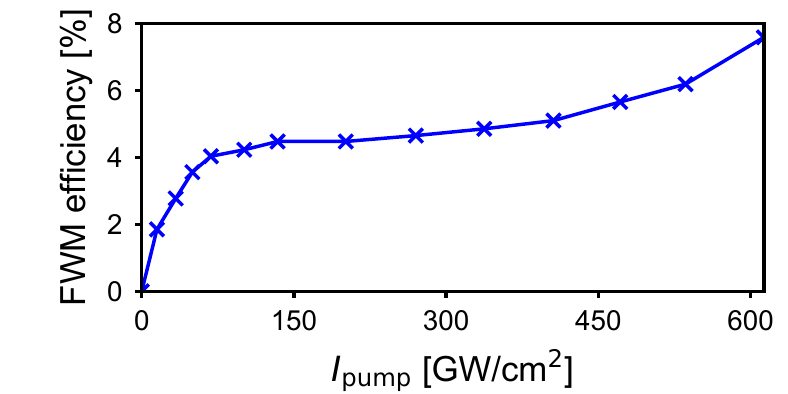}
    \caption{FWM efficiency of the sample, defined as the ratio of FWM to probe power, as a function of modulating pump intensity.}
    \label{fig:supp4}
\end{figure}

FWM signal was acquired with a Princeton Instruments NIRvana spectrometer.  Assuming this unmodulated probe signal was 7\% of the full probe power, we computed efficiency as the ratio of detected FWM to probe power, the dependence of this efficiency on the pump intensity being shown in Fig.\ref{fig:supp4}.

\section{Pump self modulation}

\begin{figure}
    \centering
    \includegraphics{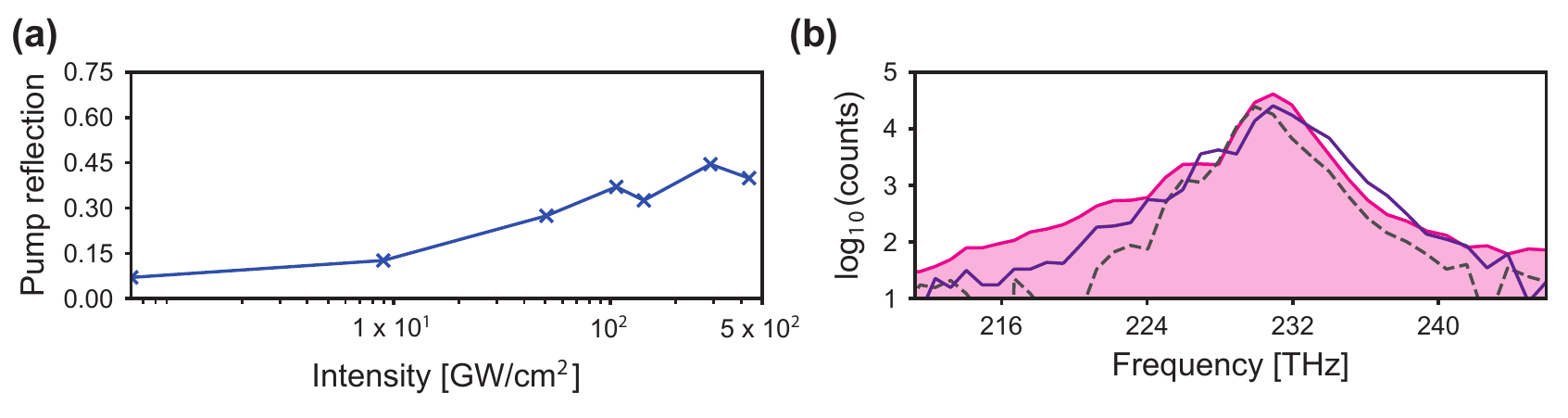}
    \caption{\textbf{(a)} Pump reflection as a function of pump intensity. \textbf{(b)} Reflected pump spectrum at high intensity (436~GW/cm$^2$, purple line) resulting from self-modulation, against original pump spectrum (dashed grey) and modulated probe spectrum (pink line) as seen in Fig.3(g).}
    \label{fig:supp5}
\end{figure}

The self-induced modulation of the pump modifies the  reflectivity of the mirror, which increases up to $R=0.45$ (Fig.\ref{fig:supp5}(a)). At the same time the pump spectrum broadens, as shown in Fig.\ref{fig:supp5}(b). The increase in reflectivity of the pump and the pump broadening is smaller than that of the probe due to the absence of matching delay of the pump with itself.



\bibliography{supplementary}